\documentclass[]{interact}

\usepackage[ruled,vlined]{algorithm2e}

\usepackage{epstopdf}
\usepackage{subfigure}

\usepackage{graphicx}

\usepackage{natbib}
\bibpunct[, ]{(}{)}{;}{a}{}{,}
\renewcommand\bibfont{\fontsize{10}{12}\selectfont}

\theoremstyle{plain}

\theoremstyle{definition}

\theoremstyle{remark}

\begin{document}


\title{Spatial Social Network (SSN) Hot Spot Detection: Scan Methods for Non-Planar Networks}

\author{
\name{Joshua Baker\textsuperscript{a}, Clio Andris\textsuperscript{a}\thanks{CONTACT Clio Andris, School of City and Regional Planning, School of Interactive Computing, Georgia Institute of Technology. Email: clio@gatech.edu}, and Daniel DellaPosta\textsuperscript{b}}
\affil{\textsuperscript{a}Georgia Institute of Technology, Atlanta, GA, USA; \textsuperscript{b}Pennsylvania State University, University Park, PA, USA}
}

\maketitle

\begin{abstract}

Moving window and hot spot detection analyzes are statistical methods used to analyze point patterns within a given area. Such methods have been used to successfully detect clusters of point events such as car thefts or incidences of cancer. Yet, these methods do not account for the connections between individual events, such as social ties within aneighborhood. This paper presents two GIS methods, EdgeScan and NDScan, for capturing areas with high and low levels of local social connections. Both methods are moving window processes that count the number of edges and network density, respectively, in a given focal area (window area). The focal window attaches resultant EdgeScan and NDScan statistics to nodes at the center of the focal window area. 

We implement these methods on a case study of 1960s connections between members of the Mafia in New York City. We use various definitions of a focal neighborhood including Euclidean, Manhattan and K Nearest Neighbor (KNN) definitions. We find that KNN tends to overstate the values of local networks, and that there is more variation in outcome values for nodes on the periphery of the study area. We find that, location-wise, EdgeScan and NDScan hot spots differ from traditional spatial hot spots in the study area. These methods can be extended to future studies that detect local triads and motifs, which can capture the local network structure in more detail. 

\end{abstract}


\section{Introduction}
In point pattern analysis, the goal of hot spot detection is to find statistically-significant concentrations of an event or phenomenon. Traditional hot spot detection (e.g. the Getis-Ord GI* statistic or Ripley's K‐function) are longstanding methods that are used to detect statistically-significant localized events. These methods are often used to find clusters of incidences of cancer, locations of similar tree species, or clusters of restaurants across a city.

The purpose of this study is to create a new statistical method for detecting not just clusters of points (hot spots), but clusters of connected points (local networks) for non-planar networks (e.g. telecommunications flows, migrants, social networks). The goal is to capture areas where sets of points are \textit{both} clustered and connected. As such, we do not just presume that nearby events are related, but we use data to show they are related. For example, given only a point list of individuals who belong to a specific religious institution, we can infer whether these individuals cluster, whether they are in certain neighborhoods and how far they live from the institution. If we are also given data on ties between members, we can examine where ties tend to be local, and where these local relationships occur. Other example applications include: species that signal to each other instead of just existing in a place with similar ecological properties; individuals who not only live nearby, but also call one another; or organizations that actually exchange ideas and technologies, in addition to simply locating in the same office park.We aim to create a statistic that assesses the significance of these non-planar spatial networks.

One theoretical principle behind geographical nearness is that co-location makes it easier to interact. The proposed statistics will let us know if interaction is actually being realized. If so, this invites questions such as: does nearness drive ties or do ties drive nearness? As a result of the ties, is there more power among this set of places for leveraging ownership of their own space and solidarity for negotiating needs? Or, does the place need special infrastructure for these ties, or do these ties help or threaten the local area, as with criminal activity \citep{Papachristos2013}? Simply put, it is more meaningful to use direct evidence of connectivity/interaction, than the inference that nearness leads to interaction.

We present two scan (moving window) statistics, EdgeScan and NDScan, which detect the number of non-planar edges and the network density, respectively, of a subset of a social network contained within a focal window. We apply these statistics to a case study of connections between geolocated members of the Mafia. We use Euclidean distance, Manhattan distance and K-nearest-neighbor (KNN) neighborhoods of different sizes to test the sensitivity of our results. Our results show the sensitivity of EdgeScan and NDScan to different window distinctions, and our recommendations. They also show the difference in hot spot locations of EdgeScan and NDScan vs. a traditional Ripley's K function and we discuss the implications of using each approach. 

This research aims to improve upon current computing efforts by adding a connectivity metric to typical point pattern hot spot detection. It does this by testing new moving window scan methods that are not limited to accounting for characteristics of focal points (as is the prevailing state-of-the-art) but detects whether these points are \textit{connected} as well. 

\section{Related Work}
Research on networks in geocomputation and GIS is typically conducted on hydrological or infrastructure, i.e. planar, networks \citep{batty2005network}. This research involves describing properties of the network itself, movement on the network, or using network distance to improve geospatial model. Network (or cost) distance is used instead of Euclidean distance to create a more realistic model of point dispersion and distance between points \citep{borruso2008network}. Common planar network analysis tasks include geospatial optimization and routing, facility placement and studies of infrastructure resilience, response and dynamics \citep{oliver2016spatial, hay2016transport}. 

Spatial network hot spot detection is a newer technique that has been performed on planar spatial networks \citep{okabe2006sanet, ishioka2019evaluation}. This technique detects point patterns that are relegated to a planar network (most often, the road network) by searching for sets of points that occur on or near adjacent edges. These methods improve upon hot spot detection methods that use circular neighborhoods, because they limit the potential area of point events to the network's geometry (e.g. omitting areas like a field where traffic accidents do not occur). These methods are also useful because they avoid grouping events on parallel or distant lines \citep{oliver2016spatial}.

In the geographical sciences, non-planar network edges (sometimes referred to as desire lines) are rarely considered in geocomputation \citep{Andris2018}, and spatial analysis methods are not often or easily applied to non-planar networks \citep{tao2016spatial}. The problem of integrating non-planar flows into GISystems for both analysis and computation has been noted as a major issue that makes it challenging for GIS analysts to use non-planar data \citep{rae2009spatial}. Prior work on crating heat maps of point-to-point migration flows across England has captured non-planar network density by counting the number of edges that cross a grid cell and creating a raster surface of results \citep{rae2009spatial, Rae2011}. While this approach is helpful for addressing a system in its entirety, it represents fly-over edges that may not be pertinent to research questions on local connections. 

Our research follows this lead, using a social network of personal connections instead of a large origin-destination matrix, and concentrating only on short-distance edges. This method can be used to respond to research questions on local social network behavior, such as likelihood of face-to-face meetings and interaction with the underlying geography. 

\section{Methods}

\subsection{Network Detection Moving Window (Scan) Methods}
We first describe two methods based on moving window scan techniques. The goal is to detect places where there are incidences of connected nodes. Like in scan statistics for general hot spot detection within point patterns, sets of neighborhoods are scanned systematically. Here, \textit{neighborhood} is defined using the common GIS/spatial statistics definitions, specifically: a set of $K$ events (e.g. $K$=15) is a point and its 14 closest neighbors) or a circle with a radius $r$ (e.g. a point and all other points within 1-kilometer buffer from the point) \citep{Kulldorff1995SpatialDisease}. In raster computing, a neighborhood is also defined as a specific number of cells, but since networks are not typically represented as rasters, we do not consider this definition here \citep{Mu2019Neighborhoods}.

\subsubsection{EdgeScan} EdgeScan is a moving window technique that surveys a network node by node. It uses a predefined neighborhood definition (e.g. 1-mile radius buffer), and marks each focal node with the number of edges that are found (entirely) in its immediate neighborhood. That is, the number of edges that are contained within this neighborhood. This produces nodes with values that detect concentrated sets of nearby edges as opposed to just concentrated sets of points. 

The EdgeScan method moves across the area of study and calculates the number of edges that have both endpoints within a focal area $A$ (see Algorithm \ref{alg:edge}). $A$ is determined on the search window neighborhood definition used. Here, we use Euclidean distance, Manhattan distance, or the set of K nodes nearest to the focal node.\\




\begin{algorithm}[H]
\caption{EdgeScan}
\label{alg:edge}
\SetKwProg{generate}{Function \emph{generate}}{}{end}
Input: N set of nodes, E set of edges, (One of: R radius of search window, M Manhattan distance of search window, or K number of nearest neighbors in search window)\\
Map heat=new Map(Node, integer)\\
     \ForAll{Node n in N}{
        \If{Using radial or Manhattan distance window}{
          Obtain subset N' of nodes within R or M distance of n, respectively.
        }
        \If{Using K-Nearest method of window}{
        Obtain subset N' of K nodes within closest Euclidean distance of n.
        }
        heat(n) = \{\# of edges with both endpoints in the window.\}
      }
      return heat
\end{algorithm}

\subsubsection{NDScan} NDScan stands for network density scan. This method searches for groups of well connected nodes within a neighborhood (see Algorithm \ref{alg:nd}). It computes the \textit{network density} of this isolated set of nodes and intra-node edges within the given neighborhood. If the focal area has higher network density $D$ than the entire network, then we consider that focal region to be specially-connected.

Network density, $D$, is calculated as the ratio of the number of connections in a network to the potential connections, given the same number of nodes.

\begin{equation}
D = \frac{2e_A}{n_A(n_A - 1)}
\end{equation}\\

Where $e_A$ and $n_A$ are the numbers of edges and nodes, respectively, in region $A$. \\






\begin{algorithm}[H]
\caption{NDScan}
\label{alg:nd}
\SetKwProg{generate}{Function \emph{generate}}{}{end}
Input: N set of nodes, E set of edges, (One of: R radius of search window, M Manhattan distance of search window, or K number of nearest neighbors in search window)\\
Map heat=new Map(Node, integer)\\
   \ForAll{Node n in N}{
        \If{Using radial or Manhattan distance window}{
          Obtain subset N' of nodes within R or M distance of n, respectively.\\
          potential = |N'|(|N'|+1)/2
        }
        \If{Using K-Nearest method of window}{
        Obtain subset N' of K nodes within closest Euclidean distance of n.\\
        potential = K(K-1)/2
        }
        numEdges = \{\# of edges between with both endpoints in the window.\}\\

        heat(n) = numEdges / potential
      }
      return heat
\end{algorithm}

\subsubsection{Neighborhood definitions} 
\begin{figure} [h]
\includegraphics[width=\linewidth]{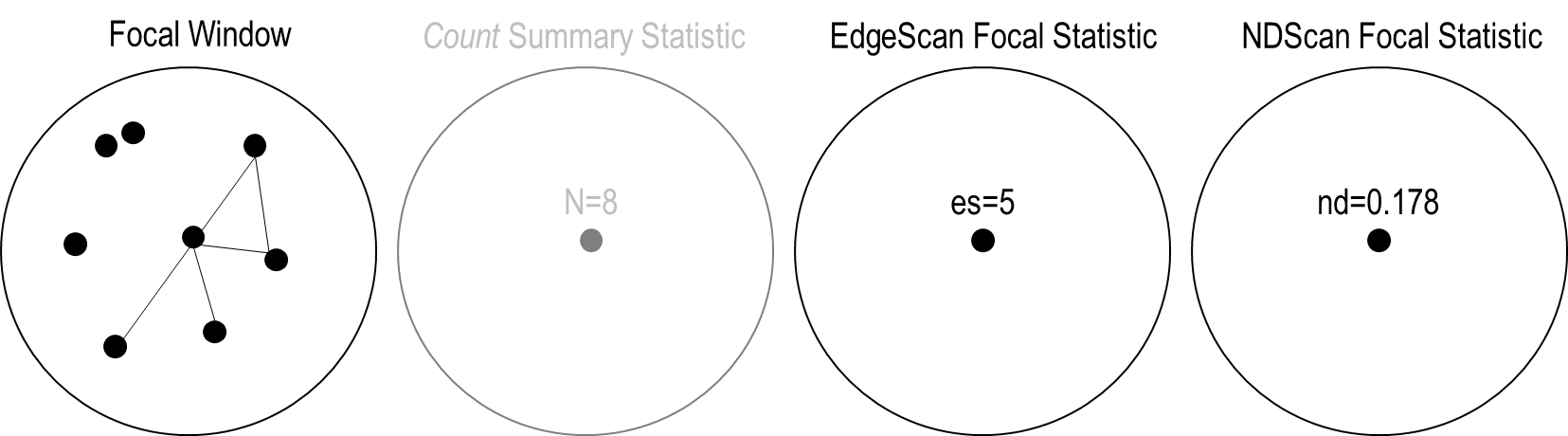}
\caption{A focal window contains eight nodes with five non-planar connections. A typical scan method would produce a summary statistic such as a `count', assigned to the central node. This research produces statistics for EdgeScan and NDScan, which account for the local network connections.}
\label{fig:focalwindows}
\end{figure}

The schematic in figure \ref{fig:focalwindows} shows the result of EdgeScan and NDScan for a nodes and links within a focal window.\\

We conduct EdgeScan and NDScan using various neighborhood definitions (Table \ref{table:neighdef}), and report on the sensitivity of the results to variation in these definitions in the results section. EdgeScan discovers hot spots where edges exist between two nodes, while NDScan accounts for both the edges and the nodes in the locality. For this reason, NDScan results can be high for very small clusters of points (such as three points), given that they are connected while EdgeScan needs a certain number of nodes in order to have a high value. Local disconnected nodes (e.g. isolates) will affect the NDScan outcome, but not the EdgeScan outcomes.

\begin{table} [h]
\tbl{Ranges for analysis neighborhood definition}
{\begin{tabular}{lc} \toprule
Neighborhood Definition&Ranges\\ 
\midrule
Euclidean Distance & 0.5, 1, 2 km radius\\
Manhattan Distance & 0.5, 1, 2 km radius \\
K Nearest Neighbors & 10, 15, 20 \\\bottomrule
\end{tabular}}
\label{table:neighdef}
\end{table}

\subsection{Granularity} Both EdgeScan and NDScan resultant statistics are tabulated and visualized at the node level. It is also possible to use the above methods to produce a statistical surface (e.g. a raster grid) of values. Many scan statistics produce such continuous raster grids, as the moving window has been traditionally used on data that represents a continuous spatial field such as elevation or rainfall. Since our input data represent discrete locations of individuals, we use the node level to help avoid uncertainty associated with estimating values in places where no individual existed. Depending on future research questions, it may be prudent to estimate results for locations where no individuals lives, if a research is interested finding, for example, that a restaurant is a proximal meeting place for a local SSN.

Regarding visualization, we view results at the node level, instead of the edge level, in order to prevent a haystack visualization problem-- a common challenge for non-planar network visualization in GISystems \citep{Andris2018}.

\subsection{Case Study Application}
We test the EdgeScan and NDscan methods on data showing connections between members of U.S. Mafia families in the 1960s (from \citep{dellaposta2017network}). Mafia families are criminal organizations whose members engage in a variety of illegal (e.g. extortion, illegal gambling, drug trafficking) and sometimes legal (e.g. some members owned bars or restaurants) economic activities or ``rackets.'' While there are many varieties of organized crime, U.S. Mafia families specifically include groups that descended from similar groups in southern Italy and restricted full membership to ethnic Italians (though non-Italians could still be ``associates'' of the family and we include such individuals). As described in qualitative and historical accounts of the U.S. Mafia \citep{Abadinsky:1983tv,Gambetta:1996wn}, a member's network of connections was perhaps their most important resource for carrying out criminal activities. For this reason, it is important to not just document clusters of co-located nodes but also the connections between them that facilitated collaboration in risky and dangerous criminal activities. 

\begin{figure}[!h]
\begin{center}
\includegraphics[width=10cm]{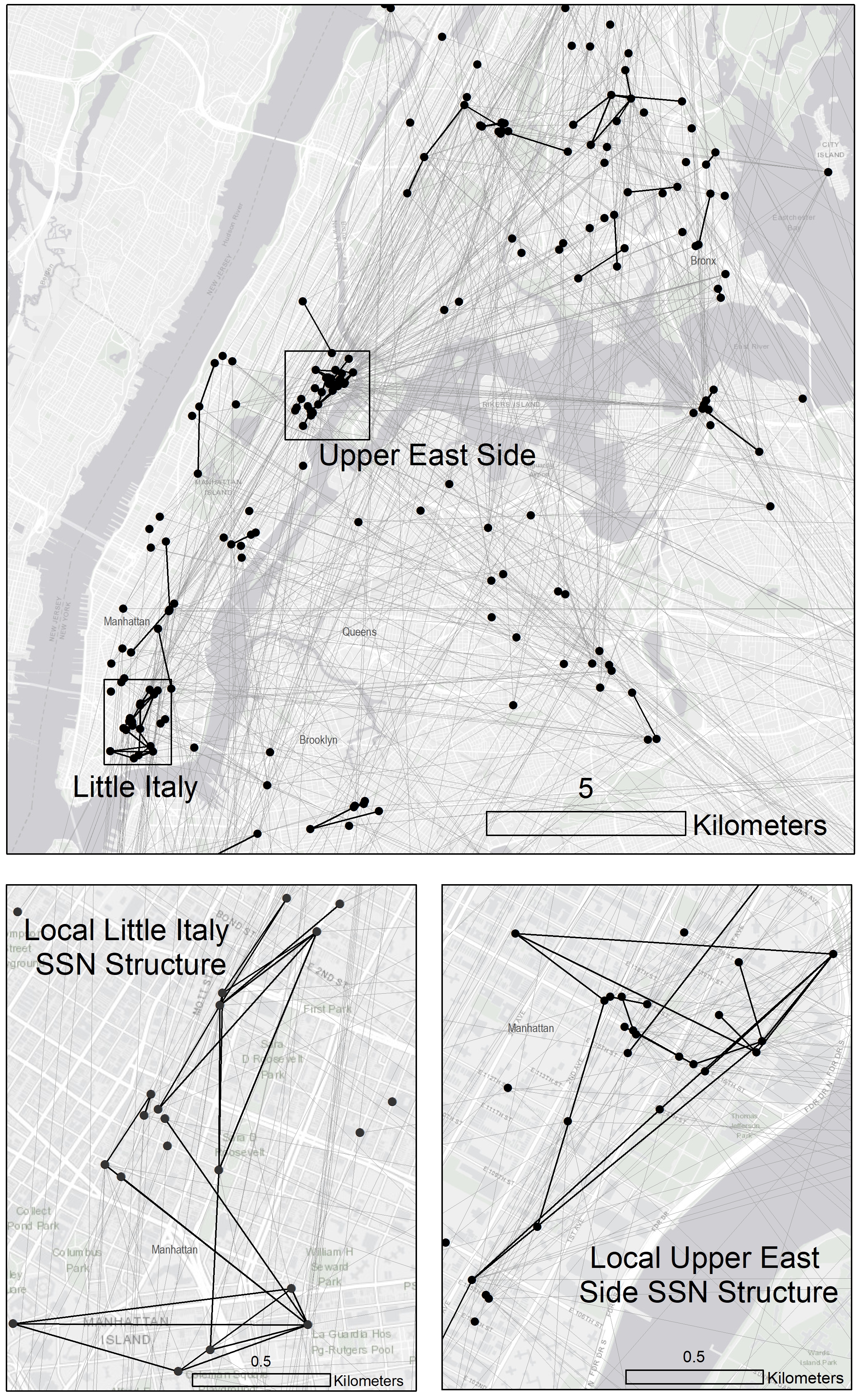}
\caption{Example of different local structures within dense clusters of points in Little Italy and the Upper East Side.}
\label{fig:mapinset}
\end{center}
\end{figure}

Our study area is the five boroughs of New York City and peripheral nearby nodes (all within New York State). We use only edges between members in this region for this case study, although members have alters outside of the study area. The data includes 298 nodes and 936 edges, a density of 0.0198 and average degree of 6.429. The average distance of an edge is 12.775 km, and the standard deviation is 10.525 (max = 55.3 km). There are four edges with 0 distance, as these individuals live at the same address (a home or apartment building). The local structure of these ties reflect particular geometries that represent a system of personal relationships between the households of individuals (Figure \ref{fig:mapinset}). Triads and isolates both emerge, as well as a few examples of 'cliques'. While we do not study the outcomes of the various structures, we seek to leverage GIS methods to capture general properties about their connectedness. It can be assumed that ties affect the stability of a sector of the city, and that this sector also has features that can host the ties. 

\begin{figure}[!h]
\begin{center}
\includegraphics[width=10cm]{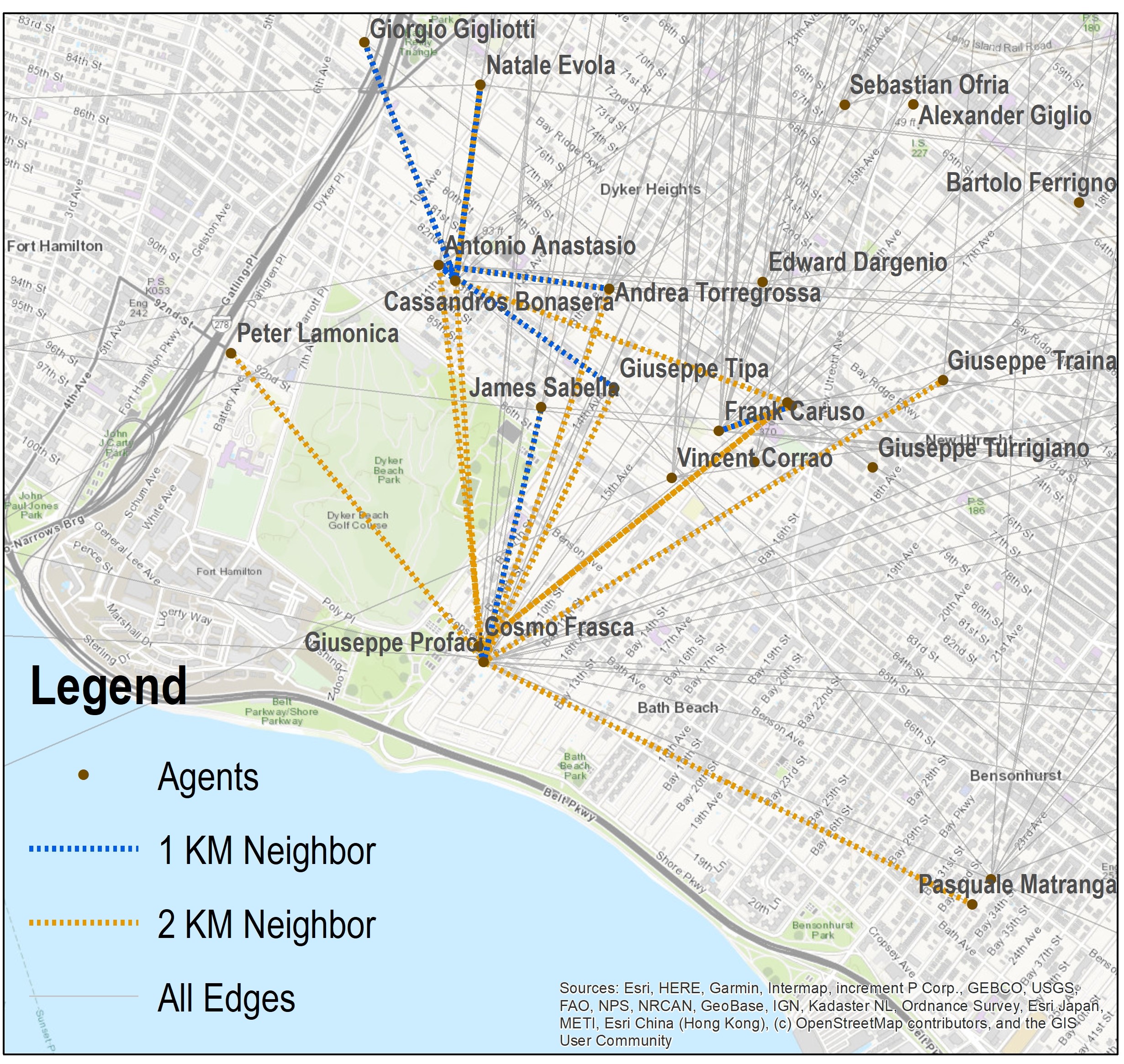}
\caption{Examples of differences in ties that would be captured using two Euclidean distance neighborhood definitions. At the 2 km level, all dotted ties (blue and orange) are counted as local edges, whereas at the 1 km level, only the blue ties are counted as local edges.}
\label{fig:method}
\end{center}
\end{figure}

An example is provided to show the difference in neighborhood definition for connected individuals of the Profaci family in Brooklyn (Figure \ref{fig:method}). This concentration of individuals has relatively few local ties, and mixed levels of distant ties (not shown).

\subsubsection{Implementation}
Code development and computation was conducted in Python. VSCode was used as a text editor, and the following Python packages were used: NumPy, Pandas, GeoPandas, descartes, Matplotlib, and geoplot. Mapping, distance calculations and hot spot detection was conducted in ArcGIS, using the New York State Plane East (NAD 83) coordinate system.


\section{Results}
We test our social network scan methods on data showing connections between members of the Mafia. First, we explore the results of EdgeScan and NDScan for different definitions of neighborhood. Next, we examine the variance of results for each individual node. Finally, we examine the difference in results between a traditional hot spot density map and the results derived from the new scan methods. 

\subsection{EdgeScan Results}
The average number of detected non-planar edges per node ranges from 0.67 to 31.17 and naturally grows with the size of the neighborhood (Table \ref{tab:EdgeNDResults}). Because EdgeScan simply detects number of edges, it is more likely to increase with increased area. The smallest values are found for 0.5 km of Manhattan distance, where nearly 75\% of nodes have no edges within this radius. The largest values are found for KNN, and are larger with more neighbors. When at least 15 neighbors are considered, no nodes have zero edges in their neighborhood. When ten neighbors are considered, only two nodes have no local edges nearby.

\begin{table}
\tbl{EdgeScan and NDScan results by neighborhood definition}
{\begin{tabular}{lcccc} \toprule
Neighborhood & Mean (St. Dev.) &N(nodes=0) (\%) & Mean (St. Dev.) &N(nodes=0) (\%) \\
\midrule
& EdgeScan & & NDScan & \\
\midrule
Euclidean (0.5 km)	&	1.12 (2.42)	&208 (69.8\%)&0.04 (0.1)	&221 (74.2\%)	\\
Euclidean (1 km)	&	3.39 (5.99)	&151 (50.7\%) &	0.06 (0.1)	&160 (53.7\%)	\\
Euclidean (2 km)	&	7.44 (9.20)	&	71 (23.8\%)	&	0.07 (0.09)	&110 (37.0\%)	\\
Manhattan (0.5 km)	&	0.67 (1.60)	&	223 (74.8\%)&0.02 (0.07)&242 (81.2\%)	\\
Manhattan (1 km)	&	2.16 (4.12)	&	169 (56.7\%)&0.06 (0.1)	&180 (60.4\%)	\\
Manhattan (2 km)	&	5.55 (7.84)	&	100 (33.6\%) & 0.07 (0.9)	&104 (34.9\%)\\
KNN (K = 10)	&	8.79 (4.44)	&2 (0.7\%) &0.16 (0.08)	&	2 (0.7\%)		\\
KNN (K = 15)	&	19.32 (7.99)	&	0 (0\%)	&0.16 (0.07)	&	0 (0\%)		\\
KNN (K = 20)	&	31.17 (11.91)	&	0 (0\%)	&0.15 (0.06)	&	0 (0\%)		 \\\bottomrule
\end{tabular}}
\label{tab:EdgeNDResults}
\end{table}

\begin{figure}[ht]
\begin{minipage}[b]{0.49\linewidth}
\centering
\includegraphics[width=\textwidth]{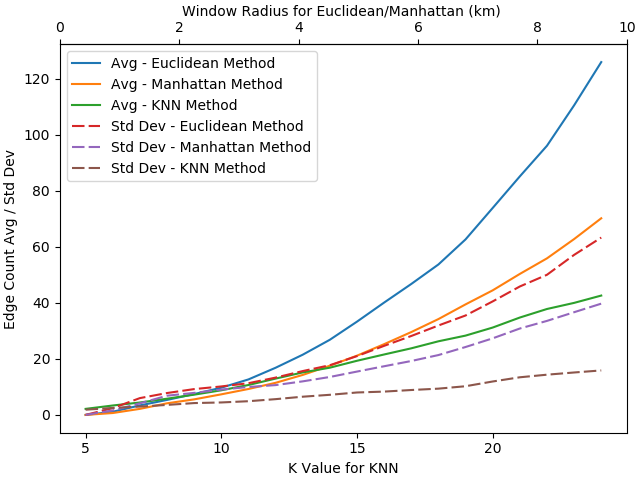}
\end{minipage}
\hspace{0.02cm}
\begin{minipage}[b]{0.49\linewidth}
\centering
\includegraphics[width=\textwidth]{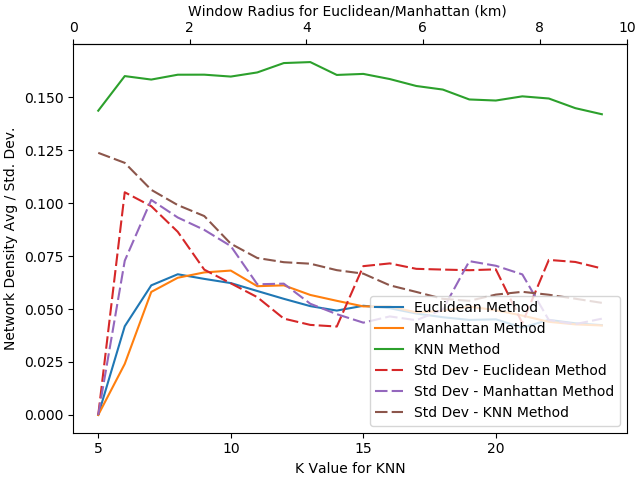}
\end{minipage}
\caption{EdgeScan and NDScan results for different neighborhood definitions. \textbf{Left:} As the windows sizes increase (dual x axes), the average count of the edges (across all 298 nodes) and variance of the values detected by EdgeScan increase. \textbf{Right:} The average density of the local nodes (across all 298 nodes) and variance (dotted lines) of the values tends to rise and plateau for different windows sizes (dual x axes).}
\label{fig:stdev}
\end{figure}

\begin{figure}[ht]
\begin{minipage}[b]{0.49\linewidth}
\centering
\includegraphics[width=\textwidth]{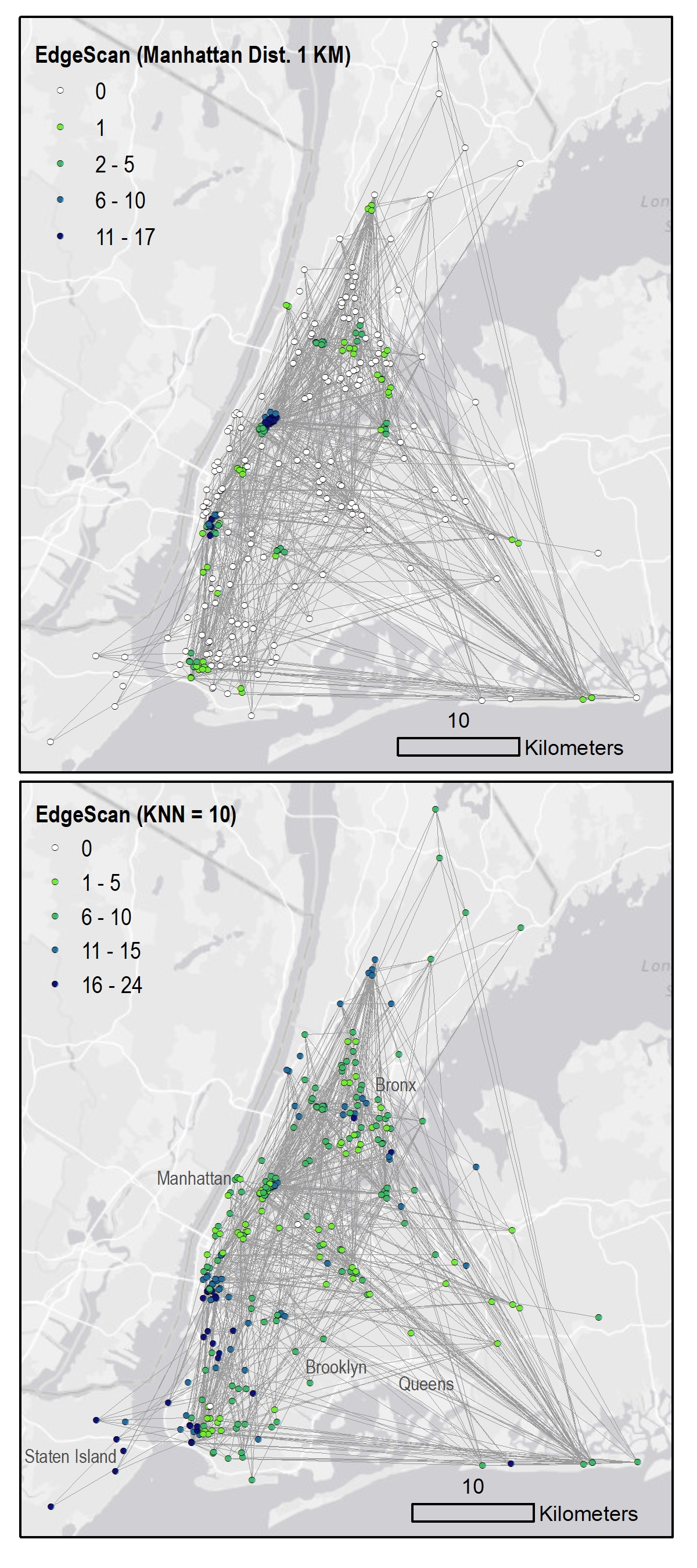}
\end{minipage}
\hspace{0.02cm}
\begin{minipage}[b]{0.49\linewidth}
\centering
\includegraphics[width=\textwidth]{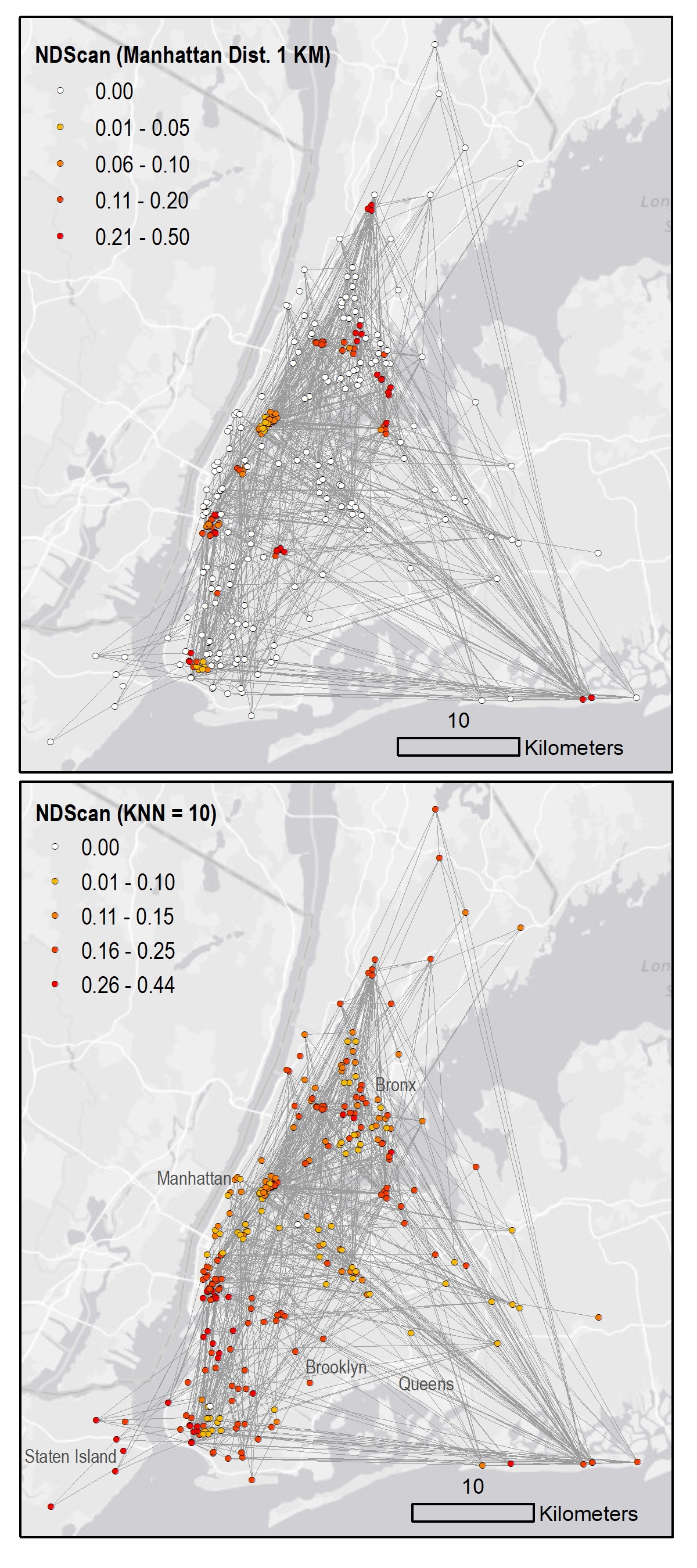}
\end{minipage}
\caption{EdgeScan and NDScan results for neighborhood definitions of: Manhattan Distance of 1 km (top) and KNN = 10 (bottom). \textbf{Left:} Each node is colored by the number of edges in its neighborhood. The top map shows more localized clustering of edges, while the bottom map shows that high edge rates are more distributed. \textbf{Right:} Each node is colored by network density of the focal nodes in its neighborhood. Like in EdgeScan results, top map also shows localized density, while the bottom map shows a dispersed sent of density values.}
\label{fig:MapEdgeND}
\end{figure}

We now examine the change in EdgeScan results across a continuous set of neighborhood definitions. Again, increases in the size of neighborhood, either by number of nearest neighbors, or radius, will have an effect on the outcome of the values. For EdgeScan, the KNN method has the smallest variance between all values as more nearest neighbors are added (Figure \ref{fig:stdev}). Using Euclidean distance, with each increment of the radius after 3 km, there is an exponential increase in the average edge count across all of the nodes. This result is expected as the area of a circle increases exponentially with linear growth of the radius. This plot shows that while K is between 5-12 and either radius is less than 3 km (for both Euclidean and Manhattan distance), the results change little with neighborhood size. Afterwards, the results of EdgeScan vary more drastically, if Euclidean distance is chosen. After about 16 neighbors or 5 km, there is also a divergence in results depending on whether Manhattan distance or KNN is used.

Prior research has asserted that 1 km is a reasonable walking distance for meeting \citep{yang2012walking}. Given this measurement, and that the average degree for nodes is roughly 6.4, we visualize scan outcomes for neighborhood of 1 km Manhattan distance and ten nearest neighbors (KNN=10). The EdgeScan methods show tight concentrations of edges in the Little Italy section of lower Manhattan, and the Upper East Side (marked with dark blue in Figure \ref{fig:MapEdgeND}). The KNN method reveals a more dispersed result, where nodes in Staten Island are counted as having high local edges, because the window joins these nodes to points in Brooklyn. As such, two different interpretations of the locations of high clusters of local relationships are given, but consensus locations do emerge.

\subsection{NDScan Results}
The average NDScan results show that regardless of type of distance definition, the local density of networks exceeds the density of the entire network (0.0198) (Table \ref{tab:EdgeNDResults}). Only the Manhattan distance neighborhood of 0.5 km has a similar network density of 0.02. However, for this neighborhood definition, over 80\% of the nodes in question have a density value of 0.The most connected networks (up to density of 0.16) occur for KNN, indicating that, regardless of distance, a lack of intervening opportunities between nodes may cause individuals to connect. In other words, ties were often made with the nearest people. This reflection holds true for 10 or 15 neighbors, but when 20 neighbors are reached, the density decreases, as these neighbors are potentially outside of the optimal range of more convenient opportunities. In fact the NDScan density of the network slightly decreases with added K neighbors, especially from about 13 - 19 neighbors (Figure \ref{fig:stdev}). However, the variance in results reveals high heteroskedasticity of NDScan values calculated using a small numbers of neighbors. This variance decreases with more neighbors, until about 17 neighbors, in which the variance plateaus, and neighborhoods with greater than 17 neighbors give similar values between nodes.

Euclidean and Manhattan distance measures for NDScan yield results that also show a plateau effect, likely due to the geography of the study area (specifically, the island of Manhattan). Yet, these results are abrupt in that from radii of 0 - 2 km, the network density increases quickly, as does the variance of this value among the nodes. However, at about 1.75 km, there is little difference in the average neighborhood density with increased area. The neighborhood will have a steady density of 0.075 regardless of size or choice to use Euclidean or Manhattan distance (Figure \ref{fig:MapEdgeND}).

Regarding the geographic results, the NDScan method reveals that the Manhattan distance neighborhood consistently yields more clustered results than the KNN results. This observation still holds, as Manhattan neighborhood distance largely depends on a fixed area radius (Figure \ref{fig:MapEdgeND}). As such, using Manhattan distance results in more localized clusters than KNN. Using Manhattan distance, the NDScan clusters resemble those of EdgeScan with a few differences. There are a few smaller clusters that are detected (marked in red), while most areas have NDScan results of 0. This shows strong peaks and troughs of the density value. It is valuable because it finds places where people are \textit{associated}, instead of simply \textit{proximal}. The NDScan results using ten nearest neighbors results in a more even surface with only two nodes with a density of 0. This method finds a number of nodes with density up to 0.33. These highly connected nodes are located mostly in Brooklyn. A few such nodes are in Manhattan and near the Bronx, and even Staten Island. While there is a longstanding ferry that connects Staten Island with Brooklyn, a better measure of cost distance in the 1960s would be needed to assess whether this grouping is a meaningful geographic cluster. However, it is notable that agents between the two boroughs of New York connected in that location. 

There are two nodes with have a value of 0 local edges and network density even when considering a neighborhood using 10 nearest neighbors (bottom maps in Figure \ref{fig:MapEdgeND}). One such node is in the center of the point pattern, perhaps indicating that it may be between two separate groups, although this node does not have connection to either side (i.e. not playing a broker role). The second node with zero values for both EdgeScan and NDScan lies north of a dense network of Profaci Family members (one of the five family organizations based in New York City) who is clustered in the southern part of Brooklyn. The zero values for this node indicates that while the node is near other Mafia members, these members are not connected. 

\subsection{Variance in Results}
\begin{figure}[ht]
\begin{center}
\includegraphics[width=9cm]{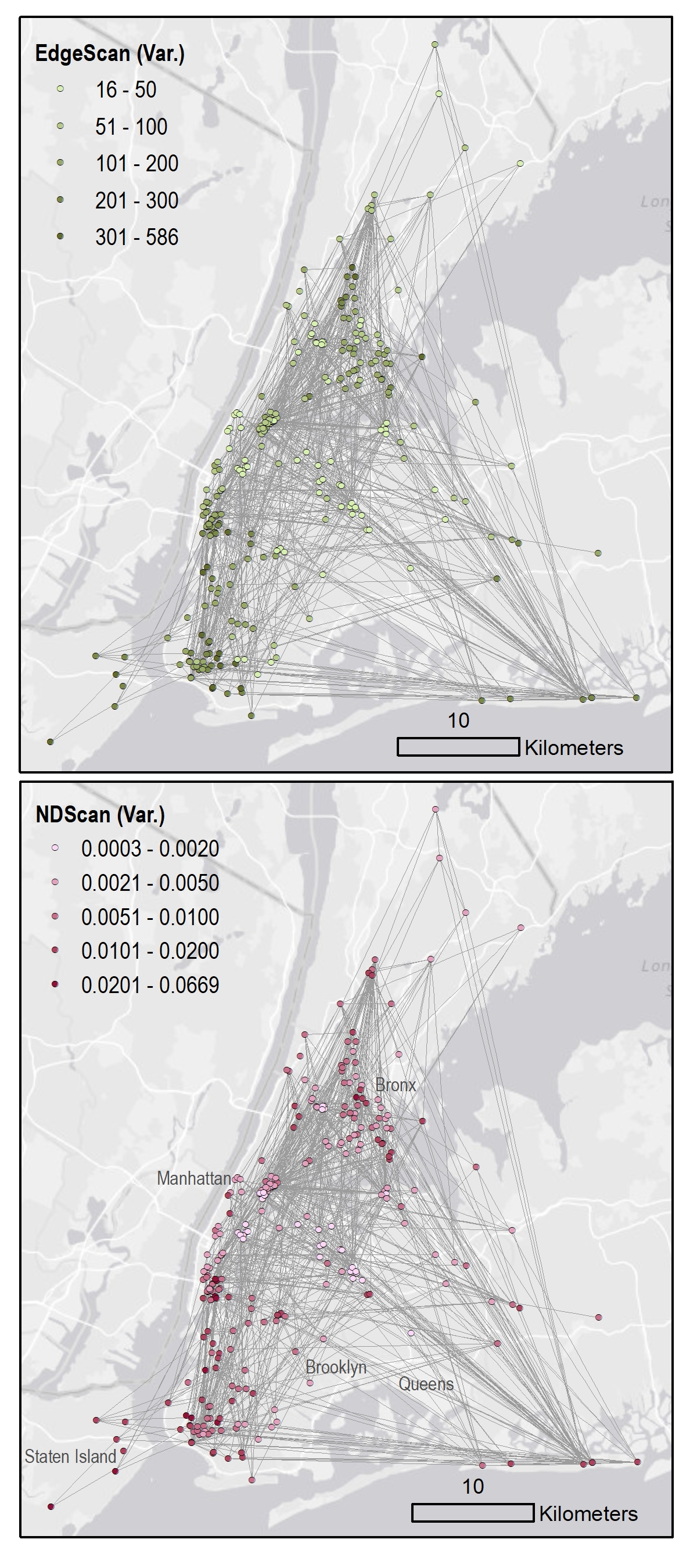}
\caption{Nodes colored by the variance of the resulting values for each node, given the nine different neighborhood distinctions shown in Table \ref{tab:EdgeNDResults}. The top image shows the variance of node values for EdgeScan and the bottom shows results for NDScan.}
\label{fig:var}
\end{center}
\end{figure}

We now examine the overall variance in results as part of our sensitivity analysis. A single value of variation is computed for each node, by measuring the variance between the values of the results from the nine neighborhood distinctions shown in Table \ref{tab:EdgeNDResults}. The variance of values for EdgeScan and NDScan are shown in Figure \ref{fig:var}. For both EdgeScan and NDScan, resultant node values tend to vary more on the periphery of the network, exhibiting a classic spatial computing problem of edge effects. Notably, for nodes with many zero values for both scan methods are likely to have a lower variance. However a number of nodes that are located in spatial clusters did not have low EdgeScan and NDScan values, and also have low variance. These maps can be used as a guide for assessing the reliability of scan method results, given a node's location.

\subsection{Triads}
We also experimented with measuring local \textit{triads} (triangles) present in the graph, and local \textit{transitivity index}, defined as the ratio of $n$(triads) to number of possible triads \citep{newmanPark2003}, given the local structure in area $A$. The Mafia SSN has 549 triads and a global transitivity index of 0.0001. For three Euclidean distance neighborhood definitions, we calculated the mean local transitivity index and mean number of triads in that neighborhood. We find that at a 2 km distance radius, we are likely to find at least one (1.28) local triangle (Table \ref{tab:TransitivityResults}).

\begin{table}
\tbl{Transitivity index results}
{\begin{tabular}{lcc} \toprule
Neighborhood & Mean (Std. Dev.) & Mean \# of Triangles \\
\midrule
Euclidean (0.5 km) & 0.0044 (0.025)	& 0.07 \\
Euclidean (1 km) & 0.0029 (0.0087)	& 0.52 \\
Euclidean (2 km) & 0.0027 (0.0086) 	& 1.37 \\ 		\bottomrule
\end{tabular}}
\label{tab:TransitivityResults}
\end{table}

\subsection{Comparison with Traditional Hot Spot Detection}
We compare our values to results from a classic node density scan in each location and compare these statistics to our actual data. We first created a surface using Getis-Ord GI* statistic function to detect statistically-significant clusters of points (with no value associated with each) using a 1 km search radius. We conducted the same test for nodes marked by NDScan results using a 1 km-Euclidean distance neighborhood definition. Results show that areas where nodes form hot spots are not the same areas as the NDScan method detects hot spots (Figure 7). There are five total grid cells marked as hot spots using the traditional point cluster method (two cells are adjacent). In only one of these cells (which has only 90\% confidence in southern Brooklyn) is there a statistically-significant NDScan cluster. There are five other adjacent clusters of nodes that are hot spots according to NDScan, and none of these intersect with the traditional results. This result suggests that there is an added benefit to detecting clusters of connected nodes, as they highlight locations where connections exist. Results for EdgeScan values (not shown) reveal (only) two major EdgeScan hot spots, each with many participating nodes. These hot spots matched the traditional hot spot areas in two concentrated locations (Upper East Side and Little Italy), and two other traditional locations did not have any EdgeScan hot spots.

\begin{figure}[ht]
\begin{center}
\includegraphics[width=10cm]{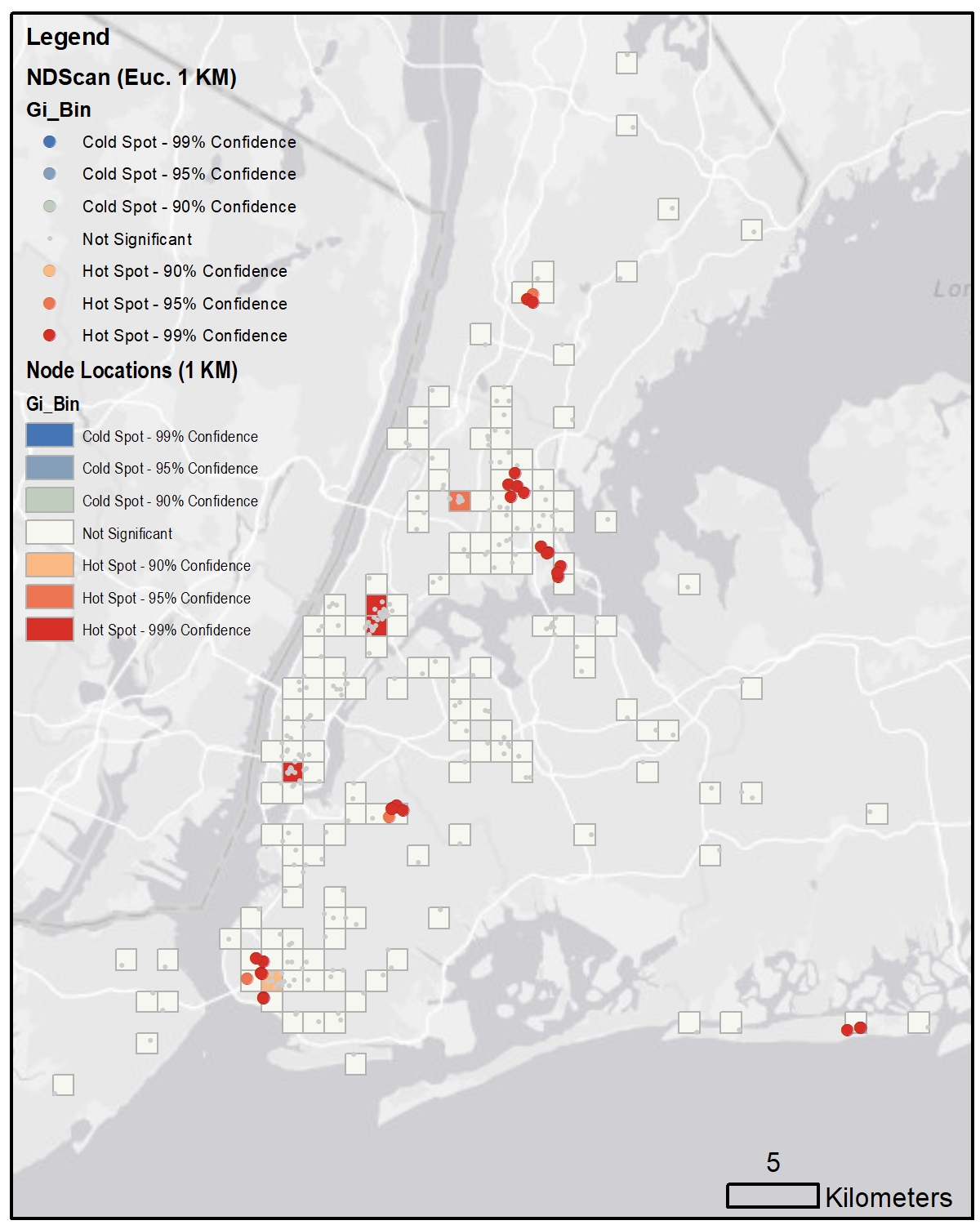}
\caption{A Getis Ord GI* hot spot detection results for typical point clusters (red grid cells) vs. NDscan results (red dots). Results show that the locations of the statistically-significant point clusters are in different locations than statistically-significant NDScan values.}
\end{center}
\label{fig:hotspot}
\end{figure} 

\section{Discussion and Future Work}
In this work, we attempted to discover localized areas where a spatial social network (SSN) has a high density of ties using two moving window methods: EdgeScan and NDScan. Using a social network of Mafia ties in the New York City area, we detected areas with a high concentration of edges and a high local network density. We tested the variance of results for different neighborhood definitions to help users choose an appropriate window size and shape for analysis. We found that using KNN tended to overestimate values, while smaller neighborhoods may have missed some important ties. Overall, there is variation in results depending on what neighborhood definition is used, and the variation tended to be more pronounced for nodes on the network's periphery. 

Regarding hot spot detection, we found that the locations of what can be considered a hot spot using NDScan and EdgeScan were different than hot spots detected by the traditional Getis Ord GI* method. This finding indicates that this traditional method may be better suited for detecting independent incidences of unconnected events. When events are connected (and these connections are known) the scan methods here present a better conceptual framework for detecting hot spots of dependent events. For our case study, these hot spots represent areas where meetings may have happened, or associates could be found walking together. They address places where coordination and local information transfer could have occurred.

A drawback to this method is that the local network structure (i.e. network motif) is not captured, only the most basic metrics of connectedness. Future work will further address the structure of the focal networks. 

Other future work includes experimenting with spatially permuting the network to create different values for ``expected" number of geographic ties in each location. We will stochastically randomly reassign our n edges to two individuals, similar to an Erdos-Renyi model (or by preserving the degree distribution of each node per the configuration model, as applied in the network flattening ratio \citep{sarkar2019metrics}). The rewiring method can be improved to include wiring probabilities that decay with distance \citep{anderson2020representing} or even attempt to maintain the distance distribution of the edges. This approach will allow us to measure the probability of our actual network configuration having formed, given the point pattern.

Future work includes `smart' determination of an optimal window size depending on how connected each node is in the graph. The goal is to consider a radial window of sufficient size wherein most of the connections to the central node are included. As an example, we consider for each node, having a search window that is large enough to encompass n\% of the incident edges to that node. Since EdgeScan and NDScan methods are dependent only on the edges incident to the central node to each search window, performing these methods on this "optimal-sized" search area would give a relevant measure to how connected that node is to its immediate surrounding region. In essence, this seeks to answer two questions: How far do we have to search in order to capture all relevant connections to each node, and once we have searched at that distance, is that node missing out on a substantial number of closer connections.

Lastly, as implemented, this scan method does not account for geographic features that may alter the actual expected values. Take, for instance, a study of relationship connections over New York City such as meeting places, parks, restaurants, churches, etc. In general, making the method more robust to capture such features help inform studies of interpersonal interaction. In addition, a more reliable measure of cost distance (including ferries and subway travel modes), beyond Manhattan distance should be implemented. 
	
\newpage

\bibliographystyle{plainnat}
\bibliography{main}

\end{document}